\documentclass[twocolumn,twocolappendix]{aastex6}

\usepackage{bm}
\usepackage{url}
\usepackage{amsmath}
\usepackage{color}
\usepackage{txfonts}

\shorttitle{}
\shortauthors{Ueda et al.}

\begin{document}

\title{
Analytic Expressions for the Inner-Rim Structure of Passively Heated Protoplanetary Disks
}


\email{t\_ueda@geo.titech.ac.jp}

\author{Takahiro Ueda\altaffilmark{1},  Satoshi Okuzumi\altaffilmark{1}, and Mario Flock\altaffilmark{2}}

\altaffiltext{1}{Department of Earth and Planetary Sciences, Tokyo Institute of Technology, Meguro, Tokyo, 152-8551, Japan}
\altaffiltext{2}{Jet Propulsion Laboratory, California Institute of Technology, Pasadena, California 91109, USA}

\begin{abstract}
We analytically derive the expressions for the structure of the inner region of protoplanetary disks based on the results from the recent hydrodynamical simulations.
The inner part of a disk can be divided into four regions: dust-free region with gas temperature in the optically thin limit, optically thin dust halo, optically thick condensation front and the classical optically thick region in order from the inside.
We derive the dust-to-gas mass ratio profile in the dust halo using the fact that partial dust condensation regulates the temperature to the dust evaporation temperature.
Beyond the dust halo, there is an optically thick condensation front where all the available silicate gas condenses out. 
The curvature of the condensation surface is determined by the condition that the surface temperature must be nearly equal to the characteristic  temperature $\sim 1200{\,\rm K}$.
We derive the mid-plane temperature in the outer two regions using the two-layer approximation with the additional heating by the condensation front for the outermost region. 
As a result, the overall temperature profile is step-like with steep gradients at the borders between the outer three regions.
The borders might act as planet traps where the inward migration of planets due to gravitational interaction with the gas disk stops.
The temperature at the border between the two outermost regions coincides with the temperature needed to activate magnetorotational instability, suggesting that the inner edge of the dead zone must lie at this border. 
The radius of the dead-zone inner edge predicted from our solution is $\sim$ 2--3 times larger than that expected from the classical optically thick temperature.
\end{abstract}


\section{Introduction}
The inner region of protoplanetary disks is the birthplace of rocky planetesimals and planets.
One preferential site of rocky planetesimal formation is the inner edge of the so-called dead zone (e.g., \citealt{Kretke2009}).
The dead zone is the location where magneto-rotational instability (MRI, \citealt{Balbus1998}) is suppressed because of poor gas ionization \citep{Gammie1996}.
In the inner region of protoplanetary disks, the dead zone is likely to have an inner edge where the gas temperature $T$ reaches $\sim$1000~K, above which thermal ionization of the gas is effective enough to activate MRI (\citealt{Gammie1996}, \citealt{Desch2015}).
Across the dead zone inner edge, the turbulent viscosity arising from MRI steeply decreases from inside out, resulting in a local maximum in the radial profile of the gas pressure (e.g., \citealt{Dzyurkevich2010}, \citealt{Flock2016}, \citealt{Flock2017}).
The pressure maximum traps solid particles (\citealt{Whipple1972}, \citealt{Adachi1976}) and may thereby facilitate dust growth (\citealt{Brauer2008}, \citealt{Testi2014}) and in-situ planet formation \citep{Tan2015}.
The dead zone edge may even trap planets by halting their inward migration \citep{Masset2006}.
Because the location of the dead zone inner edge is determined by the gas temperature, understanding the temperature structure is important to understand rocky planetesimal formation as well as the orbital architecture of inner planets.

The temperature structure of the innermost regions of protoplanetary disks has been extensively studied in the context of their observational appearance (see \citealt{Dullemond2010} for review, and also \citealt{Natta2001}; \citealt{Dullemond2001}; \citealt{Isella2005}).
\citet{Flock2016} performed the radiation hydrodynamical calculations of the  the inner rim structures of disks around Herbig Ae stars taking into account the effects of starlight and viscous heating, evaporation and condensation of silicate grains and gas opacity.
They found that the inner disk consists of four distinct zones with different temperature profiles.
According to their results, the dead zone inner edge occurs at the border between two outermost regions,
where the temperature drops steeply across 1000 K.
Thus, this complex temperature structure potentially influences where 
rocky planetesimals and planets preferentially form. 
However, there has been no simple analytic model that reproduces the entire structure of the temperature profile.
In addition, because \citet{Flock2016} performed simulations only for the disks around Herbig Ae stars, it has been unclear how the inner rim structures depend on stellar parameters. 

In this paper, we analytically derive the temperature and dust-to-gas mass ratio profile for the inner region of disks and reveal how these structures depend on the central star parameters based on the results from \citet{Flock2016}.
Our analysis and the analytic solutions are showed in Section \ref{sec:analysis}.
We discuss the implications for the planetesimal and planet formation and the disk observation in Section \ref{sec:discuss}.
The summary is in Section \ref{sec:summary}.

\section{Analytic Solutions for the Inner Rim Structure}
\label{sec:analysis}
We derive analytic expressions for the inner rim structure of protoplanetary disks based on the recent numerical simulations conducted by \citet{Flock2016}.
They found that the inner region of protoplanetary disks can be divided into four regions (region A, B, C and D) in terms of the temperature profile.
\begin{figure*}[ht]
\begin{center}
\includegraphics[scale=0.6]{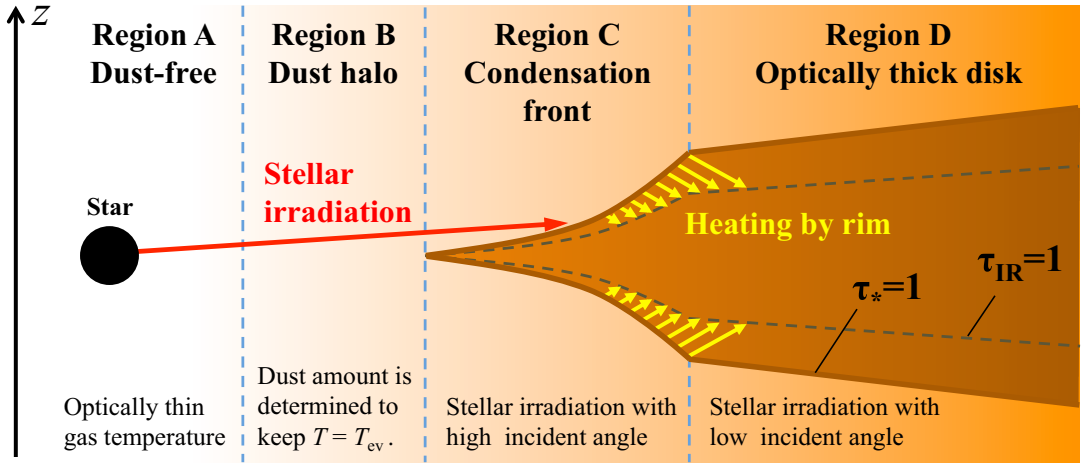}
\caption{
Schematic illustration of the inner structure of a passively irradiated protoplanetary disk. 
The disk received the starlight on the surface where the visual optical depth $\tau_*$ measured along the starlight reaches unity (solid curve), and receives the reprocessed light from the $\tau_* = 1$ surface 
on the deeper surface where the infrared optical depth $\tau_{\rm IR}$ measured vertically ($z$-direction) from above reaches unity (dashed curve).
The inner region  can be divided into four regions (region A, B, C and D).
Region A: optically thin dust-free region;
Region B: optically thin dust halo region where the temperature is nearly equal to the evaporation temperature;
Region C: optically thick dust condensation front irradiated by the central star with high incident angle; 
Region D: optically thick disk illuminated by not only the central star but also dust condensation front.
}
\label{fig:image}
\end{center}
\end{figure*}
Figure \ref{fig:image} schematically shows the inner region of protoplanetary disks.
In the following subsections, we describe how the structures such as the temperature and the dust-to-gas mass ratio are determined in each region.

\subsection{Region A: Optically thin dust-free region}
Region A is the innermost region where the temperature is above the evaporation temperature of dust. 
This region is therefore free of dust and is optically thin to the starlight.  
In an optically thin region, the temperature profile is generally given by 
 \begin{eqnarray}
T=\epsilon^{-1/4}\left( \frac{R_{*}}{2r}\right)^{1/2}T_{*}, \label{eq:regionA1}
\end{eqnarray}
where $r$ is the distance from the central star, and $R_{*}$ and $T_{*}$ are the stellar radius and temperature, respectively. The dimensionless quantity $\epsilon$ is the ratio between the emission and absorption efficiencies 
of the disk gas including the contribution from dust, 
\begin{eqnarray}
\epsilon \equiv 
\frac{\kappa_{\rm g}+f_{\rm d2g}\kappa_{\rm d}(T)}{\kappa_{\rm g}+f_{\rm d2g}\kappa_{\rm d}(T_*)}, \label{eq:regionA2}
\end{eqnarray}
where $f_{\rm d2g}$ is the dust-to-gas mass ratio, $\kappa_{\rm g}$ is the Planck mean gas opacity (here assumed to be independent of the temperature), and $\kappa_{\rm d}(T_*)$ and $\kappa_{\rm d}(T)$ are the Planck mean dust opacities at the stellar and disk temperatures, respectively.
\citet{Flock2016} adopted $\kappa_{\rm g}=10^{-4}~{\rm cm^{2}~g^{-1}}, \kappa_{\rm d}(T_*)=2100~{\rm cm^{2}~g^{-1}}$ and $\kappa_{\rm d}(T)=700~{\rm cm^{2}~g^{-1}}$.

For region A, $\epsilon \approx 1$ because $f_{\rm d2g} \approx 0$.
Therefore, the temperature profile in region A, $T_{\rm A}$, is given by
\begin{equation}
T_{\rm A}=\left( \frac{R_{*}}{2r}\right)^{1/2}T_{*}.
\label{eq:regionA3}
\end{equation}
\citet{Flock2016} showed that Equation~\eqref{eq:regionA3} 
accurately reproduces the temperature profile in region A (see the top panel of their Figure~1). 
The real gas temperature might be more difficult to understand because it depends on gas opacity (\citealt{Dullemond2010}, \citealt{Hirose2015}), which is assumed to be independent on wavelength in this paper.

\subsection{Region B: Optically thin dust halo region}
Region B is the location where dust starts to condense but is still optically thin.
This region, which \citet{Flock2016} called the dust halo, has a uniform temperature that is nearly equal to the evaporation temperature $T_{\rm ev}$ (see Figure 1 of \citet{Flock2016}).
These features can be understood by noting that the condensing dust acts as a thermostat:
if $T$ falls below $T_{\rm ev}$, dust starts to condense, but the condensed dust acts to 
push the temperature back up because the emission-to-absorption ratio $\epsilon$ of the dust 
is lower than that of the gas. 
Thus, dust evaporation and condensation regulate the temperature in this region to $\sim T_{\rm ev}$.
Strictly speaking, the temperature of region B obtained by \citet{Flock2016} is higher than 
$T_{\rm ev}$ by about 100 K. 
However, this is due to their numerical treatment of the dust evaporation, 
where the dust-to-gas ratio is given by a smoothed step function of $T$
 with the smoothing width of 100 K.

\begin{figure}[h]
\begin{center}
\includegraphics[scale=0.4]{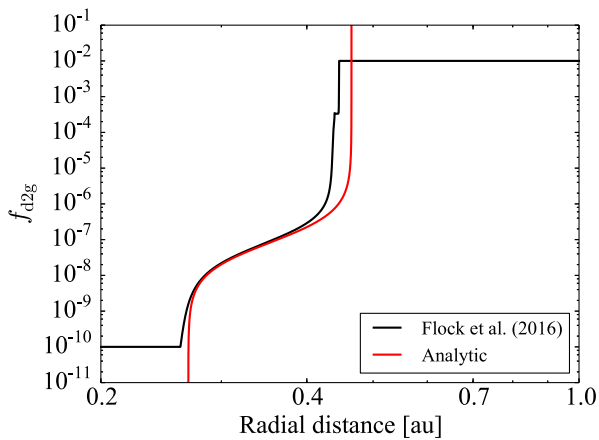}
\caption{
Dust-to-gas mass ratio profile $f_{\rm d2g}$ at the midplane in region B.
The red and black solid lines show our analytic solution (Equation \eqref{eq:regionB1}) and the radiation hydrostatic disk model S100 of \citet{Flock2016} with the highest resolution, respectively.
}
\label{fig:fd2g}
\end{center}
\end{figure}

The fact that $T\sim T_{\rm ev}$ in region B can be used to predict the spatial distribution 
of the dust-to-gas mass ratio $f_{\rm d2g}$ in this region.
Substituting Equation (\ref{eq:regionA2}) into Equation (\ref{eq:regionA1}) and solving the equation 
with respect to $f_{\rm d2g}$, we obtain
\begin{eqnarray}
f_{\rm d2g}=\frac{\kappa_{\rm g}(4r^{2}T_{\rm B}^{4}-R_{*}^{2}T_{*}^{4})}{\kappa_{\rm d}(T_*)R_{*}^{2}T_{*}^{4}-4\kappa_{\rm d}(T_{\rm B})r^{2}T_{\rm B}^{4}}, 
\label{eq:regionB1}
\end{eqnarray}
where $T_{\rm B}$ stands for the temperature in region B. Assuming that $T_{\rm B}$ is constant, 
Equation~\eqref{eq:regionB1} determines $f_{\rm d2g}$ as a function of $r$. 
Figure \ref{fig:fd2g} compares Equation \eqref{eq:regionB1} with the radial profile of $f_{\rm d2g}$ at 
the midplane taken from the radiation hydrostatic disk model S100 of \citet[][see the bottom panel of their Figure 1. 
The stellar temperature, radius, mass and luminosity are set to $T_{*}=10000\,{\rm K}$, $R_{*}=2.5R_{\odot}=0.0116\,{\rm au}$, $M_{*}=2.5M_{\odot}$ and $L_{*}=56L_{\odot}$, respectively.
See also their Table 1 for details]{Flock2016}.
Equation~\eqref{eq:regionB1} perfectly reproduces the result of \citet{Flock2016} 
when $T_{\rm B}$ is set to $T_{\rm ev}+100~{\rm K}\approx 1470~{\rm K}$, where we have used that 
$T_{\rm ev}$ in region B is $\approx 1370~{\rm K}$ in their calculation.

As we can see from Figure~\ref{fig:fd2g}, the inner and outer edges of region B 
correspond to the locations where $f_{\rm d2g}$ given by Equation \eqref{eq:regionB1}
goes to zero and infinity, respectively.  
The radii of these locations are given by
\begin{eqnarray}
R_{\rm AB} &=&\frac{1}{2} \left( \frac{T_{*}}{T_{\rm B}}\right)^{2}R_{*} \nonumber \\
&=&0.27 \left( \frac{T_{*}}{10^{4}\,\rm K} \right)^{2}
 \left( \frac{T_{\rm B}}{1470\,\rm K}\right)^{-2}
 \left(\frac{R_{*}}{2.5R_{\odot}} \right) \,{\rm au}
\label{eq:rab}
\end{eqnarray}
and
\begin{eqnarray}
R_{\rm BC}  &=& \frac{1}{2}\left( \frac{\kappa_{\rm d}(T_*)}{\kappa_{\rm d}(T_{\rm B})} \right)^{1/2}  \left( \frac{T_{*}}{T_{\rm B}}\right)^{2}R_{*}\nonumber\\
 &=& 0.46\left( \frac{\kappa_{\rm d}(T_*)/\kappa_{\rm d}(T_{\rm B})}{3} \right)^{1/2}
 \left( \frac{T_{*}}{10^{4}\,\rm K} \right)^{2} \nonumber \\
 &&\times \left( \frac{T_{\rm B}}{1470\,\rm K}\right)^{-2}
 \left(\frac{R_{*}}{2.5R_{\odot}} \right) \,{\rm au},
 \label{eq:rbc}
\end{eqnarray}
respectively. 
We have expressed the inner radius of region B as $R_{\rm AB}$
because it also stands for the outer radius of the dust-free zone (region A).
Similarly, $R_{\rm BC}$ also stands for the inner radius of the optically thick inner rim (region C). 
We note that our expression for $R_{\rm BC}$ is equivalent to that of \citet{Monnier2002}.

Equations~\eqref{eq:rab} and \eqref{eq:rbc} give the orbital radii of the boundaries of region B 
if $T_{\rm B} (\approx T_{\rm ev})$ is given. 
In fact, $T_{\rm ev}$ weakly depends on the gas density, and hence on $r$. 
Therefore, if one wants to determine $R_{\rm AB}$ and $R_{\rm BC}$ precisely, 
one has to solve Equations~\eqref{eq:rab} and \eqref{eq:rbc} simultaneously with 
the equation for $T_{\rm ev}$ as a function of $r$. This is demonstrated in the Appendix. 

As we showed, a dust halo naturally appears. 
The dust halo is a main contributor of the near-infrared emission in the spectral energy distribution of disks around Herbig Ae/Be stars (e.g., \citealt{Vinko2006}).

\subsection{Region C: Optically thick dust condensation front} \label{sec:regC}
\begin{figure}[h]
\begin{center}
\includegraphics[scale=0.4]{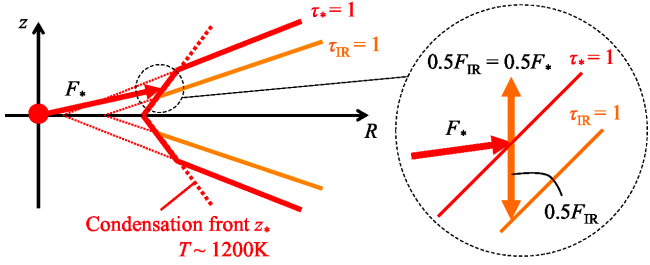}
\caption{Schematic illustration of the geometric and radiative structures of regions C and D. 
The stellar radiation $F_{*}$ determines the temperature of the optically thin layer 
above the $\tau_{*}=1$ surface.
The dust on the $\tau_{*}=1$ surface reprocesses the starlight into infrared radiation $F_{\rm IR}$, and approximately half of the dust emission 
gets absorbed by the midplane region of the disk. 
In region C, 
dust condensation/evaporation controls the location of the $\tau_{*}=1$ surface. 
}
\label{fig:front-schematic}
\end{center}
\end{figure}

At $r \approx R_{\rm BC}$, dust grains near the midplane fully condense (see Figure \ref{fig:fd2g}), 
and the visual optical depth $\tau_*$ measured along the line directly from the star
exceeds unity at the midplane. Further out, stellar irradiation is 
absorbed at a height where $\tau_{*}=1$, and the midplane temperature 
is determined by the reprocessed, infrared radiation from dust grains lying at the $\tau_{*}=1$ surface. 
Figure \ref{fig:front-schematic} schematically illustrates the surface structure of regions C and region D.
As we show below, region C can be regarded as the zone where the $\tau_{*}=1$ surface coincides with the condensation front; 
the surface structure is approximately determined by the condition that 
the surface temperature $T_{\rm s,C}$ must be equal to the $characterisctic$ temperature $\sim 1200~{\rm K}$.

\begin{figure*}[ht]
\begin{center}
\includegraphics[scale=0.85]{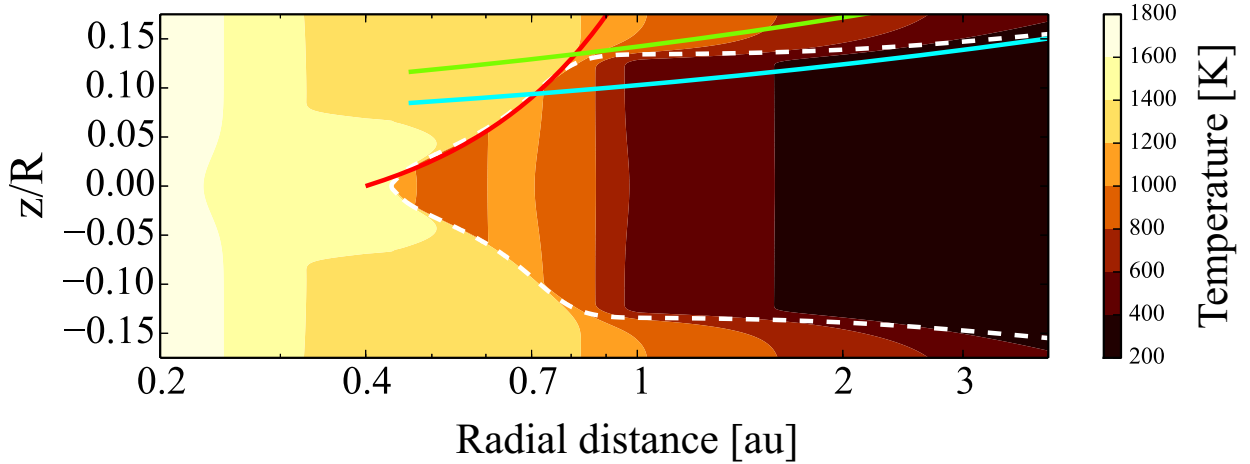}
\caption{ 
Locations of the irradiation ($\tau_{*}=1$) surface for the disk model S100 of \citet{Flock2016}
obtained from our analytic expressions (solid curves) and 
from the radiation hydrostatic calculation by \citet{Flock2016} (dashed curve).
The red solid curve shows the location of the irradiation surface in region C 
from our Equation~\eqref{eq:tau1c} with $R_{0}=0.4~{\rm au}$.
The blue and green curves show $z_{\rm *,D}=3.6h_{\rm g}$ and $4.8h_{\rm g}$, 
which well approximate the height of the irradiation surface in the inner and outer parts 
of region D, respectively. 
The curves are overlaid on the 2D temperature map obtained from the radiation hydrostatic calculation of \citet{Flock2016}.
}
\label{fig:front}
\end{center}
\end{figure*}

The radial extent of region C is of particular interest because 
the outer edge of this region sets the inner edge of the dead zone 
(\citealt{Flock2016}, \citealt{Flock2017}). We will see in Section \ref{sec:regionD} that
the border of regions C and D is related to how the height  
of the $\tau_{*}=1$ surface, $z_{*}$, depends on disk radius. 
Here, we try to estimate this height by considering the energy balance between
stellar irradiation on the $\tau_{*}=1$ surface and the thermal emission from the surface,
\begin{eqnarray}
\frac{L_{*}}{4\pi R^{2}} \sin\theta=\left( \frac{C_{\rm bw}}{4}\right)^{-1}\epsilon \sigma_{\rm SB}T_{\rm s,C}^{4}, 
\label{eq:regionC0}
\end{eqnarray}
where $R$ is the radial distance on the midplane, $\sigma_{\rm SB}$ is the Stefan-Boltzmann constant, and $\theta$ is the angle between the starlight and the disk surface (the so-called grazing angle).
$C_{\rm bw}$ is a backwarming factor (e.g., \citealt{Dullemond2001}) which is defined as the ratio of a full $4\pi$ solid angle to the solid angle subtended by the empty sky seen by particles on the surface. 
The backwarming is heating by surrounding dust particles.
If a dust particle radiates into the sky area covered by nearby dust particles, a similar amount of energy is returned from surrounding particles \citep{Kama2009}.
We assume $C_{\rm bw}=4$, which means that the $\tau_{*}=1$ surface can be approximated as a flat wall.
In region C, $\epsilon=1/3$ because dust totally condenses.
The grazing angle $\theta$ is related to the surface height $z_{*}(R)$ as (\citealt{Chiang1997}; \citealt{Tanak2005})
\begin{eqnarray}
\theta&=&\arcsin \left(\frac{4R_{*}}{3\pi R} \right) + \arctan \left( \frac{z_{*}}{R}\frac{d\ln z_{*}}{d \ln R} \right)-\arctan \left( \frac{z_{*}}{R}\right) \nonumber \\
&\approx& \frac{4R_{*}}{3\pi R} + \frac{z_{*}}{R}\frac{d\ln z_{*}}{d \ln R} -  \frac{z_{*}}{R},
\label{eq:regionC1}
\end{eqnarray}
where the second expression assumes $\theta \ll 1$.
As mentioned earlier, the $\tau_*=1$ surface in region C coincides with the condensation front, 
and the surface temperature $T_{\rm s,C}$ is approximately equal to 1200 K according to the simulations by \citet[][see their Figure 2]{Flock2016}.
Substituting Equation (\ref{eq:regionC1}) and $T_{\rm s, C} \approx {\rm constant}$ into Equation (\ref{eq:regionC0}) and solving the differential equation with respect to $z_{*}$, we obtain the surface height in region C as
\begin{eqnarray}
z_{\rm *, C}=\frac{1}{6R_{*}^{2}} \left( \frac{T_{\rm s,C}}{T_{*}} \right)^{4} (R^{3}-RR_{0}^2)+\frac{4R_{*}}{3\pi R_{0}}(R-R_{0})
\label{eq:tau1c},
\end{eqnarray}
where $R_{0}$ is the radius at which $z_{\rm *, C}=0$.
Figure \ref{fig:front} shows the location of the $\tau_{*}=1$ plane obtained by Equation \eqref{eq:tau1c} and from 3D simulation by \citet{Flock2016}.
In Figure \ref{fig:front}, we set $R_{0}=0.4\,{\rm au}$, which explains \citet{Flock2016} better than $R_{0}=R_{\rm BC}\approx 0.46 {\,\rm au}$.
Our analytic solution (Equation \eqref{eq:tau1c}) well reproduces the shape of the condensation front obtained by \citet{Flock2016}.

In order to obtain the midplane temperature in region C, $T_{\rm mid,C}$, 
we assume that the half of the infrared emission from the surface layer come into the interior.
Then we have
\begin{eqnarray}
T_{\rm mid,C}=2^{-1/4}T_{\rm s,C} \approx 1009~{\rm K}.
\label{eq:regionC}
\end{eqnarray}

\subsection{Region D: Optically thick region}
\label{sec:regionD}
Region D is the outermost, coldest region where dust condenses at all heights. 
Therefore, the temperature structure of this region is essentially the same as that of the classical two-layer disk model by \citet{Chiang1997} and \citet{Kusaka1970}. 
Specifically, the surface temperature profile in this region follows Equation (\ref{eq:regionC0})
with Equation~(\ref{eq:regionC1}), $\epsilon=1/3$, and $C_{\rm bw}=4$, i.e.,
\begin{eqnarray}
T_{\rm s,D}^{4}=3\left( \frac{R}{R_{*}} \right)^{-2} \left[ \frac{4R_{*}}{3\pi R} + \frac{z_{\rm *,D}}{R}\left(\frac{d \ln z_{\rm *,D}}{d \ln R}-1 \right) \right]T_{*}^{4},
\end{eqnarray}
where $z_{\rm *,D}$ is the surface height in region D.
Assuming that $z_{\rm *,D}$ is proportional to the gas scale height $h_{\rm g}=c_{\rm s}/\Omega_{\rm K}$, where $c_{\rm s}$ is the sound speed and $\Omega_{\rm K}$ is Keplerian angular velocity, we obtain \citep{Kusaka1970}
\begin{eqnarray}
T_{\rm s,D}^{4}=T_{1}^{4}+T_{2}^{4},
\label{eq:Td}
\end{eqnarray}
where
\begin{eqnarray}
T_{1}&=&\left( \frac{4}{\pi} \right)^{1/4} \left( \frac{R}{R_{*}} \right)^{-3/4}T_{*} \nonumber \\
&=&333 \left( \frac{R}{\rm 1~au} \right)^{-3/4} \left( \frac{R_{*}}{\rm 2.5R_{\odot}} \right)^{3/4} \left( \frac{T_{*}}{\rm 10^{4}~K} \right)~{\rm K}
\label{eq:T1}
\end{eqnarray}
and
\begin{eqnarray}
T_{2}&=&\left( \frac{6}{7} \right)^{2/7}\left( \frac{z_{\rm *,D}}{h_{\rm g}}\right)^{2/7} \left( \frac{R}{R_{*}} \right)^{-3/7} \left( \frac{k_{\rm B}T_{*}R_{*}}{mGM_{*}} \right)^{1/7} T_{*} \nonumber \\
&=&413 \left( \frac{z_{\rm *,D}}{h_{\rm g}}\right)^{2/7} \left( \frac{R}{\rm 1~au} \right)^{-3/7} \left( \frac{R_{*}}{\rm 2.5R_{\odot}} \right)^{4/7} \nonumber \\
&&\times \left( \frac{M_{*}}{2.5M_{\odot}} \right)^{-1/7} \left( \frac{T_{*}}{\rm 10^{4}~K} \right)^{8/7}~{\rm K},
\label{eq:T2}
\end{eqnarray}
where $k_{\rm B}$ is Boltzmann constant, $G$ is the gravitational constant and $m=4 \times 10^{-24}~{\rm g}$ is the mean molecular mass of the disk gas.
As in region C, the midplane temperature of region D is given by
\begin{eqnarray}
T_{\rm mid,D}=2^{-1/4}T_{\rm s,D}.
 \label{eq:Tdmid}
\end{eqnarray}

The border between regions C and D can be defined as the position where the $\tau_{*}=1$ surfaces in the two regions intersect.
Equating $z_{\rm *,C}$ given by Equation (\ref{eq:tau1c}) with $z_{\rm *,D} = (z_{\rm *,D}/h_{\rm g})h_{\rm g} 
= (z_{\rm *,D}/h_{\rm g}) \sqrt{k_{\rm B}T_{\rm mid,D}R^3/mGM_*}$ , 
we obtain the transcendental equation for the radius $R_{\rm CD}$
of the border, 
\begin{eqnarray}
\frac{1}{6R_{*}^{2}} \left( \frac{T_{\rm s, C}}{T_{*}} \right)^{4} (R_{\rm CD}^{2}-R_{\rm BC}^2)
+\frac{4R_{*}}{3\pi R_{\rm BC}R_{\rm CD}}(R_{\rm CD}-R_{\rm BC})\nonumber \\
=\left( \frac{z_{\rm *,D}}{h_{\rm g}} \right) \sqrt{\frac{k_{\rm B}T_{\rm mid,CD}R_{\rm CD}}{mGM_{*}}},
\label{eq:cdboundary}
\end{eqnarray}
where the subscript CD stands for the value at $R = R_{\rm CD}$.
and we have used that $R_0 \approx R_{\rm BC}$ (see Section~\ref{sec:regC}).
In order to derive a closed-form expression for $R_{\rm CD}$, we simplify Equation (\ref{eq:cdboundary}) as follows.
Firstly, we neglect the second term on the left-hand side of Equation \eqref{eq:cdboundary}
because for T-Tauri and Herbig stars the inner-rim radius is much larger than the stellar radius.
Secondly, we replace the factor $T_{\rm mid,CD}^{1/2} = (2^{-1}T_{\rm s,CD}^{4})^{1/8}$ appearing in the right-hand side of Equation \eqref{eq:cdboundary} by $(T_{2,\rm CD}^{4})^{1/8}$, where $T_{2,\rm CD}$ is 
the value of $T_2$ at $R = R_{\rm CD}$, because $T_1 \sim T_2$ at $R \sim$ 0.1--1 au 
(see Equations~\eqref{eq:T1} and \eqref{eq:T2}).
Thirdly, we neglect the $R_{\rm CD}$-dependence of the right-hand side,
 by substituting $R_{\rm CD} = 2R_{\rm BC}$ in the right-hand side only,
 because the dependence is generally weak ($\sqrt{R_{\rm CD} T_{\rm mid,CD}} \propto R_{\rm CD}^{1/8}$ 
 for $T_{\rm mid,CD} \approx 2^{-1/4}T_{1,\rm CD}$, and $\sqrt{R_{\rm CD} T_{\rm mid,CD}}  \propto R_{\rm CD}^{2/7}$ for $T_{\rm mid,CD} \approx 2^{-1/4}T_{2,\rm CD}$).
With these simplifications, Equation \eqref{eq:cdboundary} can be analytically 
solved with respect to $R_{\rm CD}$, and we thus obtain an closed-form expression for $R_{\rm CD}$, 
\begin{eqnarray}
R_{\rm CD}=\sqrt{ 1 + \Gamma } R_{\rm BC}
\label{eq:rcd}
\end{eqnarray}
where
\begin{eqnarray}
\Gamma &\equiv& 3.1\left(\frac{R_{\rm BC}}{0.46~{\rm au}}\right)^{-12/7}\left( \frac{z_{\rm *,D}/h_{\rm g}}{4.8} \right)^{8/7} \left( \frac{T_{*}}{10^{4}~\rm K} \right)^{32/7} \nonumber \\
&& \times \left( \frac{M_{*}}{2.5M_{\odot}} \right)^{-1/2} \left( \frac{R_{*}}{2.5R_{\odot}} \right)^{16/7}.
\label{eq:deltar}
\end{eqnarray}
If we use Equation~\eqref{eq:rbc} to eliminate $R_{\rm \rm BC}$ from $\Gamma$, 
we obtain
\begin{eqnarray}
\Gamma = 3.1 \left( \frac{z_{\rm *,D}/h_{\rm g}}{4.8} \right)^{8/7} \left( \frac{L_{*}}{56L_{\odot}} \right)^{2/7} \left( \frac{M_{*}}{2.5M_{\odot}} \right)^{-1/2},
\label{eq:Gamma}
\end{eqnarray}
here, we set $\kappa_{\rm d}(T_{*})/\kappa_{\rm d}(T_{\rm B})=1/3$ and $T_{\rm B}=1470~{\rm K}$.
The resultant equations (Equations \eqref{eq:rcd} and \eqref{eq:Gamma}) justify the validity of the approximation of $R_{\rm CD} = 2R_{\rm BC}$.
Our simplifications introduce an error of less than $~10$\% in the estimate of $R_{\rm CD}$; 
if we exactly solve Equation \eqref{eq:cdboundary}, we obtain $R_{\rm CD}=0.85 {\,\rm au}$ with $T_{*}=10000{\,\rm K}$, $R_{*}=2.5R_{\odot}$, $M_{*}=2.5M_{\odot}$ and $z_{\rm *,D}/h_{\rm g}=4.8$, while Equation \eqref{eq:rcd} leads to $R_{\rm CD}=0.92 {\,\rm au}$.
As seen in Figure \ref{fig:front}, the ratio $z_{\rm *,D}/h_{\rm g}$ is a function of the distance from the central star:
The ratio $z_{\rm *,D}/h_{\rm g}$ is about 3.6 at outer part and 4.8 at inner part of region D due to the extra heating by the hot rim surface discussed below.
Therefore, when we estimate the location of $R_{\rm CD}$, it is better to use the higher value for $z_{\rm *,D}/h_{\rm g}$ ($\approx 4.8$).
On the other hand, when we estimate the temperature in region D, it is better to use the lower value for $z_{\rm *,D}/h_{\rm g}$ ($\approx 3.6$).
\begin{figure}[h]
\begin{center}
\includegraphics[scale=0.4]{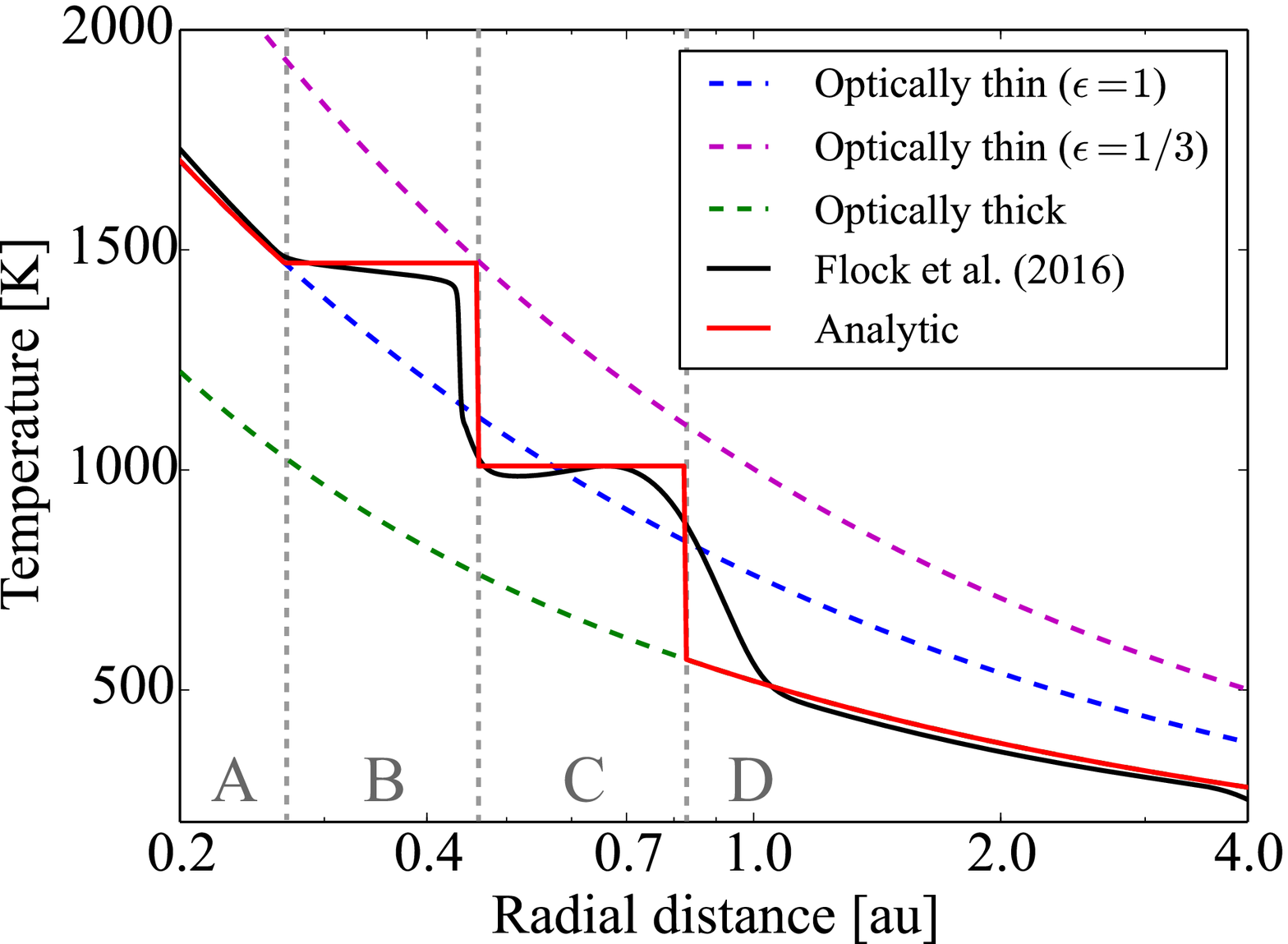}
\caption{
Radial profile of the midplane temperature $T_{\rm mid}$ 
for the disk model S100 of \citet{Flock2016}.
The black solid line is from the radiative hydrostatic calculation by \citet[][their Figure 1]{Flock2016}, 
whereas the red solid line is from our analytic expressions: 
Equation~\eqref{eq:regionA3} for region A, 
$T=T_{\rm eva}=1470~{\rm K}$ for region B,
Equation~\eqref{eq:regionC} for region C, 
Equation~\eqref{eq:Tdmid} with $z_{\rm *,D}/h_{\rm g}=3.6$ for region D.
Boundaries are determined by Equations \eqref{eq:rab}, \eqref{eq:rbc} and \eqref{eq:rcd}.
For reference, the temperature profiles for optically thin disks with $\epsilon=1$ and $1/3$ are 
shown by the blue and purple dashed lines, respectively, and the profile for a smooth optically thick disk
(Equation~\eqref{eq:Tdmid}) is shown by the green dashed line.
}
\label{fig:tempall}
\end{center}
\end{figure}

In Figure \ref{fig:tempall}, we compare our analytic temperature profile with the profile 
derived from the radiative hydrostatic disk model S100 of \citet{Flock2016}.
We find that our analytic profile well reproduces the result of \citet{Flock2016} except in the inner part of region D ($R \approx$ 0.8--1 au), where our Equation \eqref{eq:Tdmid} underestimates the midplane temperature.
This implies that some sources other than the central star heat the inner part of region D.

\begin{figure}[h]
\begin{center}
\includegraphics[scale=0.6]{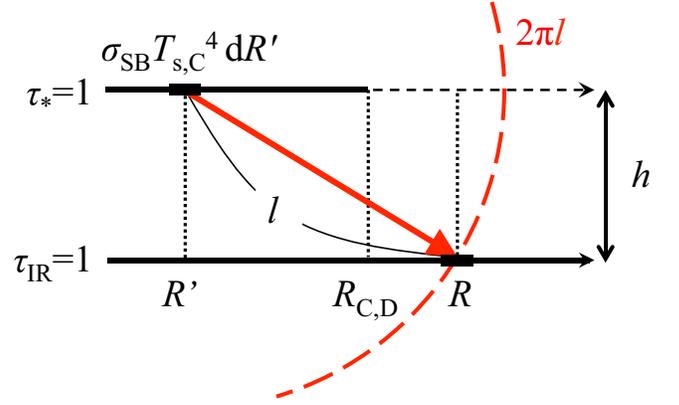}
\caption{Schematic illustration of the irradiation by the condensation front at the vicinity of the boundary between regions C and D.
A line element with a width $dR'$ at a position $R'$ on the truncated $\tau_{*}=1$ surface irradiates in the infrared wavelength with temperature $T_{\rm s,C}$.
The emission travels along the distance $l$ and gets absorbed by the position $R$ on the $\tau_{\rm IR}=1$ surface located at distance $h$ below the $\tau_{*}=1$ surface.
}
\label{fig:rcd}
\end{center}
\end{figure}

As we show below, this discrepancy can be resolved by 
taking into account the irradiation by the hot rim surface located in region C as shown in Figure \ref{fig:image}.
For simplicity, we approximate the $\tau_{*}=1$ surface of region C by a flat surface truncated at $R=R_{\rm CD}$, and assume that the $\tau_{\rm IR}=1$ surface is located at distance $h$ below the $\tau_{*}=1$ surface (see Figure \ref{fig:rcd}).
At any radial position $R$ on the $\tau_{\rm IR}$ surface, the balance between radiative cooling on the $\tau_{\rm IR}=1$ surface and irradiation heating by the hot rim surface can be written as
\begin{eqnarray}
\sigma_{\rm SB}T_{\rm mid, D}^{4}=\int_{-\infty}^{R_{\rm CD}} \frac{\sigma_{\rm SB} T_{\rm s,C}^{4}}{2\pi l}\frac{h}{l}dR'
\end{eqnarray}
with $l=\sqrt{h^{2}+(R-R')^{2}}$, where we have used that in the absence of internal heat sources, the temperature at the $\tau_{\rm IR}=1$ is equal to the midplane surface. 
Performing the integration, we obtain
\begin{eqnarray}
T_{\rm mid, D} = \left\{ \frac{1}{2\pi}\left[ \tan^{-1}\left( \frac{R_{\rm CD}-R}{h}\right) +\frac{\pi}{2} \right] \right\}^{1/4} T_{\rm s,C}.
\label{eq:tempcd}
\end{eqnarray}

Strictly speaking, the $\tau_{*}=1$ surface of region C is not parallel to the $\tau_{\rm IR}=1$ surface of region D.
From Equations \eqref{eq:tau1c} and \eqref{eq:Tdmid}, the inclinations of $\tau_{*}=1$ surfaces in regions C and D with respect to the mid-plane at $R=R_{\rm CD}$ can be estimated as $30^{\circ}$ and $10^{\circ}$, respectively.
Hence, if we assume that $\tau_{\rm IR}=1$ surfaces in regions C and D are parallel to the $\tau_{*}=1$ surfaces in regions C and D, respectively, the angle made by the $\tau_{*}=1$ surface in region C and the $\tau_{\rm IR}=1$ surface in region D is $\approx 20^{\circ}$.
This deviation from the plane-parallel approximation introduces an error of $\sim 20\%$ in temperature.
Nevertheless, our analytic formula (Equation \eqref{eq:tempcd}) based on the approximation well reproduces the result of \citet{Flock2016} with an error much less than $\sim 20\%$.
This might be because the actual position of $\tau_{\rm IR}=1$ surface in region D is not parallel to the $\tau_{*}=1$ surface in region D and should be determined by the optical depth along the ray emitted by each position on the $\tau_{*}=1$ surface in region C.
If one needs more accurate temperature structure, detailed radiative transfer simulations would be needed.

\begin{figure}[h]
\begin{center}
\includegraphics[scale=0.4]{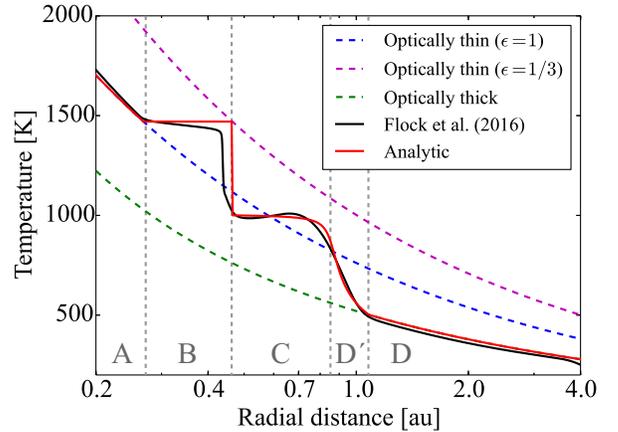}
\caption{
Same as Figure~\ref{fig:tempall} except that the 
analytic temperature profile (red line) now uses Equation (\ref{eq:tempcd}) with $h=0.05R_{\rm CD}$ 
for the midplane temperature in inner part of region D, region ${\rm D'}$.
We set $R_{\rm CD}=0.86~{\rm au}$ which explains \citet{Flock2016} better than that estimated from Equation \eqref{eq:rcd} ($0.83\, {\rm au}$ with $z_{\rm *,D}/h_{\rm g}=3.6$).
}
\label{fig:tempmod}
\end{center}
\end{figure}
Figure \ref{fig:tempmod} shows the analytic temperature profile refined with Equation \eqref{eq:tempcd}.
We set $h=0.05R_{\rm CD}$ which is a bit larger than the typical gas scale height and $R_{\rm CD}=0.86\,{\rm au}$ which explains \citet{Flock2016} better than that estimated from Equation \eqref{eq:rcd} ($\approx 0.83\, {\rm au}$ with $z_{\rm *,D}/h_{\rm g}=3.6$).
The larger value of $h$ leads to the larger radial extent of the region where the additional heating by the hot rim is important (region ${\rm D'}$)
We here use Equation (\ref{eq:tempcd}) or (\ref{eq:Tdmid}), whichever is greater, for the temperature in region D.
We find that the refined model well reproduces the temperature profile of \citet{Flock2016} in the vicinity of the boundary between regions C and D. 

\section{Discussion}
\label{sec:discuss}

\subsection{The position of the dead zone inner edge}
\label{sec:dead}
One of the motivation of this work is to determine the position of the dead zone inner edge, $R_{\rm DIB}$.
Our results show that at the boundary between regions C and D, the temperature steeply decreases from $\sim 1000~{\rm K}$ to $\sim 600 ~{\rm K}$.
According to \citet{Desch2015}, the position of the dead zone inner edge would be located at where $T\sim 1000~{\rm K}$.
Therefore, the boundary between regions C and D is thought to be the dead zone inner edge.
From Equation (\ref{eq:rcd}), $R_{\rm DIB}$ can be estimated as $0.09\,{\rm au}$ with $T_{*}=5000~{\rm K}$ and $M_{*}=1M_{\odot}$ 
and $0.8\,{\rm au}$ with $T_{*}=10000~{\rm K}$ and $M_{*}=2.5M_{\odot}$ 
with $z_{\rm *,D}/h_{\rm g}=4.8$, $\kappa_{\rm d}(T_{*})/\kappa_{\rm d}(T_{\rm B})=1/3$ and $T_{\rm B}=1470~{\rm K}$,
while the classical optically thick temperature profile (Equation~\eqref{eq:Tdmid}) leads to $R_{\rm DIB}\approx0.03\,{\rm au}$ with $T_{*}=5000~{\rm K}$ and $M_{*}=1M_{\odot}$ 
and $R_{\rm DIB}\approx0.3\,{\rm au}$ with $T_{*}=10000~{\rm K}$ and $M_{*}=2.5M_{\odot}$. 
Therefore, the location of the dead zone inner edge estimated from our model is 2--3 times farther out than that estimated from the classical optically thick temperature profile.
This is because the temperature in region C is much higher than that estimated from the classical model due to the stellar irradiation on the condensation front with high incident angle.

\subsection{The effect of viscous heating}
We have focused on passively irradiated disks where the effect of viscous heating 
on the temperature structure is negligibly small.
In the opposite case where viscous heating dominates over stellar irradiation, 
the midplane temperature $T_{\rm mid,vis}$ is given by 
\begin{eqnarray}
T_{\rm mid,vis}=\left( 1+\frac{3\tau_{\rm IR}}{4} \right)^{1/4}T_{\rm s,vis}
\label{eq:Tvis} 
\end{eqnarray}
where $\tau_{\rm IR}$ is the infrared optical depth at the midplane, and 
\begin{eqnarray}
T_{\rm s,vis}&=&\left( \frac{3GM_{*}\dot{M}}{8\pi \sigma_{\rm SB} R^{3}} \right)^{1/4} \nonumber \\
&\approx&107 \left( \frac{R}{1~\rm au} \right)^{-3/4} \left( \frac{M_{*}}{2.5M_{\odot}} \right)^{1/4}\left( \frac{\dot{M}}{10^{-8}M_{\odot}/{\rm yr}} \right)^{1/4}~{\rm K}
\end{eqnarray}
is the surface temperature determined by the mass accretion rate $\dot{M}=3\pi\Sigma_{\rm g}\nu$ \citep[e.g.,][]{Shakura}, where $\nu$ is the turbulent viscosity.   
To see when viscous heating can be neglected, 
we compare Equation \eqref{eq:Tvis} with Equation \eqref{eq:Tdmid}, which is the minimum estimate for the midplane temperature of a passively irradiated disk (see Figure \ref{fig:tempmod}). 
Combining Equations (\ref{eq:Tdmid}) and (\ref{eq:Tvis}), together with 
$\tau_{\rm IR} = (1/2)\kappa_{\rm d}\Sigma_{\rm d} = (1/2)f_{\rm d2g}\kappa_{\rm d}\Sigma_{\rm g}$
 (where $\Sigma_{\rm g}$ and $\Sigma_{\rm d}$ are the surface densities of gas and dust),  
we find that $T_{\rm mid,D} > T_{\rm mid,vis}$ if
\begin{eqnarray}
&&0.8 \left( \frac{L_{*}}{56L_{\odot}} \right)^{-2/7} \left( \frac{M_{*}}{2.5M_{\odot}} \right)^{31/70} \left( \frac{\kappa_{\rm d}}{700~\rm cm^{2}~g^{-1}} \right)^{1/5} \left( \frac{f_{\rm d2g}}{10^{-2}} \right)^{1/5} \nonumber \\
&&\times \left( \frac{\dot{M}}{10^{-8}M_{\odot}~{\rm yr^{-1}}} \right)^{2/5}  \left( \frac{\alpha}{10^{-3}} \right)^{-1/5} \left( \frac{R}{1~\rm au} \right)^{-33/70} <1.
\label{eq:viscon}
\end{eqnarray}
Here, we have used the $\alpha$-prescription for the turbulent viscosity \citep{Shakura}, i.e., $\nu=\alpha c_{\rm s }h_{\rm g}$.
We have also assumed $\tau_{\rm IR} \gg 1$, $z_{\rm *,D}/h_{\rm g}=3.6$ and $T_{\rm s,D}^{4}=2T_{2}^{4}$ for simplicity.
Equation \eqref{eq:viscon} suggests that in a disk around a Herbig Ae star of $L_* = 56L_\sun$ and $M_* = 2.5M_\sun$, and $\dot{M} = 10^{-8} M_\sun~{\rm yr}^{-1}$, stellar irradiation dominates over viscous heating if $f_{\rm d2g} \la 10^{-2}$, in agreement with the result of the hydrodynamical simulation by \citet[][see their Figure 8]{Flock2016}.
Even in a inner part $(\sim 0.1~{\rm au})$ of a disk around a T Tauri star of $L_* = L_\sun$ and $M_* = M_\sun$, stellar irradiation dominates if $f_{\rm d2g} \la 10^{-3}$ and $\dot{M} \la 6 \times 10^{-10}M_{\odot}~{\rm yr^{-1}}$.
Therefore, our temperature model is applicable to weakly accreting T Tauri disks where tiny, opacity-dominating dust is depleted (through, e.g., dust coagulation).

\subsection{Migration trap}
A steep temperature drop as seen in the borders of regions B, C, and D is known to act as a planet trap where the strong corotation torque halts the inward migration of planets due to the negative Lindblad torque from the gas disk (e.g., \citealt{Paardekooper2010}).
In addition, if the boundary between regions C and D sets the dead-zone inner edge as discussed in Section \ref{sec:dead}, the positive surface density gradient arising from the change in turbulent viscosity also prevents planets from inward migration \citep{Masset2006}.

It is meaningful to compare the location of the borders to the orbital distribution of observed planets in oder to reveal whether the borders really act as the migration trap.
Our solutions show that the positions of the boundaries between regions B, C and D are roughly proportional to $L_{*}^{1/2}$, 
though the position of the boundary between regions C and D has the additional factor $\sqrt{1+\Gamma}$ which slightly depends on the stellar parameters.green
If we assume $L_{*} \propto M_{*}^{3/2}$ \citep{Mulders2015}, these boundaries are scaled as $M_{*}^{3/4}$.
\citet{Mulders2015} investigated Kepler Objects of Interest (KOIs) and showed that the distance from the star where the planet occurrence rate drops scales with semi-major axis as $M_{*}^{1/3}$, which is different from our estimate.
However, the spectral range is not enough for the detailed comparison because the majority of the host star for KOIs are F, G and K-type stars.
In order to discuss the boundaries really act as a migration trap, observations of planets around A and B-type stars are needed.
\citet{Mulders2015} also showed that many KOIs orbit at $\sim 0.1{\,\rm au}$ which is similar to the value of $R_{\rm DIB}$ estimated from our model.

\subsection{Instabilities at the inner region of protoplanetary disks}
\label{sec:instability}
Steep temperature gradients at the boundaries between regions B, C and D would induce various kind of instabilities.
One possible instability is the subcritical baroclinic instability,  which is the instability powered by radial buoyancy due to the unstable entropy gradient (\citealt{Klahr2003}, \citealt{Lyra2014}).
If we assume the power-law density and temperature profile as $\rho \propto R^{-\beta_{\rho}}$ and $T \propto R^{-\beta_{T}}$, respectively, the condition for the subcritical baroclinic instability is given by \citep{Klahr2003} 
\begin{eqnarray}
\beta_{T} -(\gamma_{\rm ad}-1)\beta_{\rho}>0,
\end{eqnarray}
where $\gamma_{\rm ad}$ is the adiabatic index.
This condition would be satisfied at the boundaries between regions B, C and D, where $\beta_T$ is large.
The boundary between regions C and D might also have a positive surface density gradient ($\beta_\rho < 0$) if the dead-zone edge inner edge lies there. The positive density gradient would further enhance the instability on the boundary.
However, the subcritical baroclinic instability does not been observed in \citet{Flock2016}.
This is because the spatial resolution \citet{Flock2016} has is not enough to resolve the instability.
\citet{Lyra2014} showed that for resolutions $> 32$ cells/$h_{\rm g}$ the subcritical instability converges, while \citet{Flock2016} uses around 15 cells/$h_{\rm g}$.
And also, \citet{Flock2016} calculated for around 20 local orbits at $2\,{\rm au}$, which is not enough for the subcritical instability to operate.
The growth timescale for the instability requires many hundreds of local orbits \citep{Lyra2014}.

The Rossby wave instability could also be induced at the boundary between regions C and D due to the steep surface density gradient produced by the dead-zone edge \citep{Lyra2012}. 
In fact, the magnetohydrodynamical simulation by \citet{Flock2017} shows that the boundary between regions C and D produces a vortex, indicating that the Rossby wave instability should operate there.

If such instabilities operate on these boundaries, vortices caused by the instabilities would accumulate dust (e.g., \citealt{Barge1995}), which in turn could lead to rocky planetesimal formation via the streaming instability (e.g., \citealt{Youdin2005}). We will explore this possibility in future work.

\section{Summary}
\label{sec:summary}
We have analytically derived the temperature and dust-to-gas mass ratio profile for the inner region of protoplanetary disks based on the results from the recent hydrodynamical simulations conducted by \citet{Flock2016}.
The temperature profile for the inner region of protoplanetary disks can be divided into four regions.
The innermost region is dust-free and optically thin with the temperature determined by the gas opacity (Equation \eqref{eq:regionA3}).
As the temperature goes down and approaches the dust evaporation temperature, silicate dust starts to condense, producing an optically thin dust halo with a nearly constant temperature regulated by partial dust condensation.
We have derived the dust-to-gas mass ratio profile in the dust halo using the fact that partial dust condensation regulates the temperature to the dust evaporation temperature (Equation \eqref{eq:regionB1}).
Beyond the dust halo, there is an optically thick condensation front where all the available silicate gas condenses out.
The curvature of the condensation surface is simply determined by the condition that the surface temperature must be nearly equal to the characteristic temperature $\sim 1200{\, \rm K}$ (Equation \eqref{eq:tau1c}).
The temperature profile for the outermost region is essentially same as the classical optically thick temperature profile (e.g., \citealt{Kusaka1970}).
We have derived the mid-plane temperature in the outer two regions using the two-layer approximation with the additional heating by the condensation front for the outermost region.
As a result, the overall temperature profile follows a step-like profile with steep temperature gradients at the borders between the outer three regions. 
The borders might act as planet traps where the inward migration of planets due to gravitational interaction with the gas disk stops.
The temperature at the border between the two outermost regions coincides with the temperature needed to activate magnetorotational instability, suggesting that the inner edge of the dead zone must lie at this border.
The radius of the dead-zone inner edge predicted from our temperature profile is $\sim$ 2--3 times larger than that expected from the classical optically thick temperature.

\acknowledgments
We thank Neal Turner and Shigeru Ida for insightful discussions. 
We also thank the anonymous referee for useful comments on our manuscript.
S.O. is supported by Grants-in-Aid for Scientific Research (No.~15H02065,~16H04081,~16K17661).

\appendix

\section{Effect of the evaporation temperature}
Dust evaporation temperature $T_{\rm ev}$ is modeled as a function of the gas density $\rho$ as \citep{Isella2005} 
\begin{eqnarray}
T_{\rm ev}=2000\left( \frac{\rho}{1~{\rm g~cm^{-3}}}\right)^{0.0195}~{\rm K}.
\label{eq:Tev}
\end{eqnarray}
Using the relation $\rho=\Sigma_{\rm g}/\sqrt{\pi}h_{\rm g}$ and substituting Equation (\ref{eq:Tev}) as $T_{\rm B}$, Equations (\ref{eq:rab}) and (\ref{eq:rbc}) can be rewritten as 
\begin{eqnarray}
R_{\rm AB}&=&\left[ 0.079 \left( \frac{R_{*}}{0.01~\rm au}\right) \left( \frac{T_{*}}{5000~{\rm K}}\right)^{2} \right. \nonumber \\
&\times&\left. \left( \frac{\Sigma_{0}}{100~\rm g~cm^{-2}} \right)^{-0.0386} \left( \frac{M_{*}}{M_{\odot}} \right)^{-0.0193}   \right]^{\gamma} ~{\rm au} \label{eq:rab2}
\end{eqnarray}
and
\begin{eqnarray}
R_{\rm BC}=\left( \frac{\kappa_{\rm d}(T_*)}{\kappa_{\rm d}(T)} \right)^{\gamma/2}R_{\rm AB},
\label{eq:rbcev}
\end{eqnarray}
respectively.
Here, $\Sigma_{0}$ is the surface density at {1~\rm au}.
The index $\gamma$ is related to the index of the radial surface density profile, $\Sigma_{\rm g}\propto r^{-\beta}$, as $\gamma= 2.0195/(1.9025-0.078\beta)$.
For $\beta=1$, $\gamma\approx1.107$.
Equation (\ref{eq:rab2}) is similar to the expression derived by \citet[][their Equation (A.4)]{Kama2009} 
although the $\beta$ dependence is different.
If the disk is massive ($\Sigma_{0} \gg 1000{\rm ~g~cm^{-2}}$), the boundary between region B and C (i.e. the radial position for $\tau_{*}=1$) moves closer to the star compared to $R_{\rm BC}$ written by Equations (\ref{eq:rbc}) and (\ref{eq:rbcev}) because of the large optical depth.
In this case, in oder to determine the radial position of $\tau_{*}=1$, we have to calculate the optical depth using Equation (\ref{eq:regionB1}).

\bibliographystyle{./aasjournal}
\bibliography{./innerrim}

\end{document}